\documentstyle{article}

\newcommand{\bibit}{\em}
\newcommand{\be}{\begin{equation}}
\newcommand{\ee}{\end{equation}}
\newcommand{\bear}{\begin{eqnarray}}
\newcommand{\ear}{\end{eqnarray}}
\newcommand{\sH}{{\widetilde{\cal H}}}
\newcommand{\sL}{{\widetilde{\cal L}}}

\begin{document}
\begin{center}
{{\large \bf
UNIVERSAL HIDDEN SUPERSYMMETRY IN CLASSICAL MECHANICS \\ \vspace{2mm}
                AND ITS LOCAL EXTENSION\footnote{{\it This work is dedicated
to the memory of Dmitrij Vasil'evich Volkov and to all those who are 
struggling to have their work recognized. I wish to warmly thank the organizers
of the Volkov Memorial for the heroic effort of organizing it during a freezing
winter in Ukraine.}}}}\\

\vspace{3mm}
E. \, Gozzi\\
\vspace{1mm}
{\it Dept. of Theor. Physics, University of Trieste \\
         Strada Costiera 11, Miramare-Grignano \\
               34014 Trieste, Italy \\
              INFN Sezione di Trieste \\
               gozzi@trieste.infn.it }
\end{center}
\vspace{1mm}
\abstract{We review here a path-integral approach to {\it classical} mechanics
and explore the geometrical meaning of this construction. In particular we
bring to light a universal hidden BRS invariance and its geometrical relevance
for the Cartan calculus on symplectic manifolds.
Together with this BRS invariance we also show the presence of a
universal hidden genuine non-relativistic {\it supersymmetry}. In an attempt
to understand its geometry we make this susy local following the analogous
construction done for the supersymmetric quantum mechanics of Witten.}
\vskip 1cm

\section {Introduction}
The discovery of supersymmetry~\cite{su71}, first at the pure
algebraic level then  in two dimensions and finally in four dimensions
both in its  non-linear  and  linear versions, has been one of 
the most fascinating discovery of the last 30 years. Besides
softening the ultraviolet behaviour of theories of fermions coupled
with bosons, supersymmetry  offered a way out, in its unbroken phase, 
to the cosmological constant problem. 

Unfortunately the spectrum of particles 
known to-day indicates that
supersymmetry must be broken. This brings back the cosmological constant 
problem unless the breaking of this {\it special} symmetry manages to avoid the 
usual theorems on symmetry breaking. In order to explore that possibility
it is crucial to understand supersymmetry at a deeper level. 

Supersymmetry, soon after its discovery, was made
local~\cite{sugra74} producing theories of gravity that raised the hopes 
of not being afflicted by the problem of the non-removable
infinities which plagued Einstein gravity. 
Despite the huge amount of work done on supergravity and later on 
superstrings, still the goal of a finite or renormalizable theory of gravity
seems not to be near. 

All of the above mentioned issues seem to indicate that supersymmetry
is very important but its roots should be
understood better. It is our opinion that these roots are
deeply {\it geometrical}.

In this respect
there has been a 1-D model, proposed by Witten in 
1981~\cite{wi81}, which has been a toy model for exploring several issues
connected both with supersymmetry breaking and with geometry~\cite{wi82}.
The model has the limit of being just a model reproducing no aspect whatsoever
of the natural world, except some phenomenological stochastic dynamics
~\cite{pa82}. Quite independently of that model, the present author,
together with M.Reuter and W.D.Thacker, has tried~\cite{go86} to give a 
path-integral representation to Classical Mechanics (CM) and the outcome
has been a superlagrangian which has both a physical and geometrical meaning
and {\it it is not just a toy model}. So we feel that some geometrical
issues regarding supersymmetry could be tackled better using our
Lagrangian. In fact it turns out that the 
1-D superhamiltonian produced by this path-integral is nothing else than the 
{\it Lie-derivative} of the Hamiltonian flow and it  has  both a 
{\it universal} BRS-AntiBRS-like invariances
and also a {\it universal} genuine non-relativistic supersymmetry~\cite{go90}. 
The BRS-AntiBRS
invariances have been interpreted geometrically as being the exterior
derivative and co-derivative on symplectic spaces, and all the standard
Cartan calculus can be reproduced via our formalism.  As we said before
there is also a genuine supersymmetry whose physical meaning
has been studied in~\cite{go90} as being related to the 
ergodicity-non-ergodicity of the Hamiltonian system under consideration.
That supersymmetry~\cite{go90} had also other 
nice features. For example, due to the non-positivity of the Hilbert space, 
it could be broken without lifting the vacuum from zero and this can be
very important for the cosmological constant problem. Being this susy universal
it could be extracted not only from a point particle dynamics
but also from any classical field theory once they
are formulated via a Lie-brackets kind of formalism. Of course it would not
be a relativistic supersymmetry even for relativistic field theories
because it gives a privileged role to the forth component of space-time.
It might be  possible to made it relativistic if one uses the 
De Donder-Weyl~\cite{ka83} formulation of field-theory, but this has still
to be explored in details~\cite{go98}. Of course we would obtain in that case a susy at the
Lie-bracket level in which each boson and fermion present in the
basic theory, whose Hamiltonian will  act as a superpotential for our
superhamiltonian, will get a partner and produce in that manner a huge
pletora of fields which will live in a sort of shadow world degenerate
with our world. The meaning of these partners would be purely geometrical
as it is clear from~\cite{go86}. It would be interesting if these
shadow world, made only of auxiliary geometrical fields, had to be
included in the counting needed for the cosmological constant~\cite{go98}.

To better understand the geometry behind this supersymmetry, in this
brief note we take a first step in that direction
by gauging it in  the simple case of the point particle dynamics.

\section{Path-Integral for Classical Mechanics}

This section is meant to be a very quick and incomplete review of the
path-integral approach to CM. For more details the reader should
study~\cite{go86}. The "propagator" which gives
the {\it classical} probability for a particle to be at the point 
~$\phi_{2}$~at time ~$t_{2}$, if it it was at the point~$\phi_{1}$~at
time $t_{1}$, is just a delta function~$P\bigl(\phi_{2},t_{2}\vert
\phi_{1},t_{1}\bigr)={\delta}^{2n}\bigl(\phi_{2}-\Phi_{cl}(t_{2},
\phi_{1})\bigr)$~where ~$\Phi_{cl}(t,\phi_{0})$~ is a solution of 
Hamilton's equation~${\dot\phi}^{a}(t)=\omega^{ab}\partial_{b}H(\phi(t))$
subject to the initial conditions ~$\phi^{a}(t_{1})=\phi_{1}^{a}$.
Here ~$H$~ is the conventional Hamiltonian of a dynamical system
defined on some phase-space ~${\cal{M}}_{2n}$ with local coordinates~
$\phi^{a},a=1\cdots 2n$~ and constant symplectic structure~
$\omega={1\over 2}\omega_{ab}d\phi^{a}\wedge d\phi^{b}$.
Slicing in infinitesimal parts the time interval ~$[t_{2}-t_{1}]$~above 
and doing standard manipulations~\cite{go86}~, which for brevity we do not 
reproduce here, it is possible to give a path-integral representation 
to the transition probability above:
\begin{equation}
P\bigl(\phi_{2},t_{2}\vert
\phi_{1},t_{1}\bigr)=\int_{\phi_{1}}^{\phi_{2}}
{\cal D}\phi~{\cal D}\lambda~
{\cal D}c~{\cal D}{\bar c}~exp~i\int_{t_{1}}^{t_{2}}{\widetilde{\cal L}}
\end{equation}
where
~${\widetilde{\cal L}}\equiv
\lambda_{a}\bigl[{\dot\phi}^{a}-\omega^{ab}\partial_{b}H(\phi)\bigr]+
i{\bar c}_{a}\bigl(\delta^{a}_{b}\partial_{t}-\partial_{b}
[\omega^{ac}\partial_{c}H(\phi)]\bigr)c^{b}$.~The ~$\lambda_{a}$~ are 
c-number auxiliary variables while the~$c^{a}$,
${\bar c}_{a}$~ are Grassmannian auxiliary variables, so the new space
is an enlargement of the phase-space from~$2n$~dimensions to~$8n$.
The standard equations of motion for~$\phi^{a}$~ can be obtained from
the variation with respect to~$\lambda_{a}$~ of the above Lagrangian.
They can also be obtained by the introduction
of the the associated Hamiltonian
~${\widetilde{\cal L}}=\lambda_{a}{\dot{\phi}}^{a}+i{\bar c}_{a}{\dot c}
^{b}-{\widetilde{\cal H}}$, with~${\widetilde{\cal H}}$~ given by
~${\widetilde{\cal H}}=\lambda_{a}h^{a}+i{\bar c}_{a}\partial_{b}h^{a}
c^{b}$~ where ~$h^{a}(\phi)\equiv\omega^{ab}\partial_{b}H(\phi)$~are the 
components of what is called the the Hamiltonian vector field~\cite{ma78}.
To get the eqs. of motion via~$\sH$~one needs also an extended Poisson bracket 
structure~$(epb)$~in the above mentioned enlarged space. It is easy to see that they are
given by~$\{\phi^{a},\lambda_{b}\}_{epb}=\delta^{a}_{b},~\{c^{a},
{\bar c}_{b}\}_{epb}=-i\delta^{a}_{b}~,~all~ others=0$~
Note that these are different from the normal Poisson brackets
on~${\cal M}$~ which were ~$\{\phi^{a},\phi^{b}\}_{pb}=\omega^{ab}$.
Via the extended Poisson brackets above we obtain from 
${\widetilde{\cal H}}$~ the same equations of motion as those which one
would obtain (at least for ~$\phi^{a}$) from ~$H$~via the normal Poisson 
brackets: $\{\phi^a,H\}_{pb}=\{\phi^{a},{\widetilde{\cal H}}\}_{epb}$.
Regarding the ~$c^{a}$~one can easily see~\cite{go86} that they 
can be identified with the forms~$c^{a}=d\phi^{a}$, so the coordinates
~$(\phi^{a},c^{a})$~can be thought as labelling the cotangent bundle
~$T^{\ast}{\cal M}$~to phase-space. It is also easy~\cite{go86}, via the
operatorial formulation of CM, to realize that the $\lambda_{a},{\bar c}_{a}$
fields are instead a basis of the tangent bundle to the previous space.
So the ~$8n$~variables~$(\phi^{a},c^{a},\lambda_{a}.{\bar c}_{a})$~
are a set of coordinates for the tangent bundle to the cotangent
bundle to phase-space~$T(T^{\ast}{\cal M})$. This gives a complete 
{\it geometrical} description of all the various auxiliary variables 
which we had to introduce and makes this {\it not just a model} but something
more fundamental. It is also possible~\cite{go86} to make a correspondence
between higher forms and polynomials in ~$c^{a}$~and between
antisymmetric multivector fields\cite{ma78} and polynomials in
~${\bar c}_{a}$. We will indicate this
correspondence via a ~$\widehat{(\cdot)}$ symbol:
$F^{(p)}={1\over p!}F_{a_{1}\cdot a_{p}}d\phi^{a_{1}}\wedge\cdots
\wedge d\phi^{a_{p}} \Longrightarrow  {\widehat F}^{(p)}={1\over p!}
F_{a_{1}\cdots a_{p}}c^{a_{1}}\cdots\wedge c^{a_{p}}$,\linebreak
$v^{(p)}={1\over p!}V^{a_{1}\cdots a_{p}}\partial_{a_{1}}\wedge\cdots\wedge
\partial_{a_{p}} \Longrightarrow {\widehat V}^{(p)}={1\over p!}
{\bar c}_{a_{1}}\cdots {\bar c}_{a_{p}}$.
Before proceeding further we should also point out that the ~
${\widetilde{\cal H}}$~presents some universal invariance whose charges are
the following~\cite{go86}:
$Q \equiv ic^{a}\lambda_{a}$, ${\bar Q} \equiv i{\bar c}_{a}\omega^{ab}
\lambda_{b},~~
Q_{g} \equiv c^{a}{\bar c}_{a},~~
K \equiv {1\over 2}\omega_{ab}c^{a}c^{b},~~
{\bar K} \equiv {1\over 2}\omega^{ab}{\bar c}_{a}{\bar c}_{b}$.
Using the correspondence indicated above it is then possible to rewrite all the
normal operations of the Cartan calculus\cite{ma78}, like doing an exterior
derivative on forms ~$dF$~, or doing an interior product between a vector field
and a form~$i_{v}F$~, or building the Lie-derivative of a vector field~$l_{h}$
, by just using polynomials in ~$c$ and ~${\bar c}$~ together with the 
extended Poisson brackets structure and the charges built above. 
These rules, which we called
~$\{\cdot,\cdot\}_{epb}$-rules, are summarized here:
$dF^{(p)} \Longrightarrow  i\bigl\{Q,{\widehat F}\bigr\}_{epb}$~,~
$i_{v}F^{(p)} \Longrightarrow  i\bigl\{{\widehat v}, {\widehat F^{(p)}}
\bigr\}_{epb}$~;~
$l_{h}F=di_{h}F+i_{h}dF \Longrightarrow  -\bigl\{{\widetilde{\cal H}},
{\widehat F}\bigr\}_{epb}$~,~				
$pF^{(p)}  \Longrightarrow  i\{Q_{g},{\widehat F}\}_{epb}$~,~
$\omega(v,\cdot)\equiv v^{\flat} \Longrightarrow  i\{{\bar K},{\widehat V}
\}_{epb}$~,~
$(df)^{\sharp} \Longrightarrow  i\{{\bar Q},f\}_{epb}$
where the last three operations indicated  above are, respectively,
multiplying a form ~$F^{(p)}$~ by its degree ~$p$, mapping a vector field
~$V$~ into its associated one form~$V^{\flat}$~ via the symplectic form,
and building the associated Hamiltonian vector field ~$(df)^{\sharp}$~out
of a function ~$f$. One sees from above that the various abstract derivations 
of the Cartan calculus are all implemented by  some charges acting via the 
epb-brackets. From the third relation above one also can notice that the 
Lie-derivative of the Hamiltonian vector field of time evolution
becomes  nothing else than our~${\widetilde{\cal H}}$, thus
confirming that the weight-function of our classical path-integral,
generated by just a simple Dirac delta, is  the right {\it geometrical} object
associated to the time-evolution.
\section{Universal Supersymmetry}
The universal charges indicated 
in the previous section,${Q}$,${\bar Q}$,$Q_{g}$,$K$,${\bar K}$, 
have all a  geometrical meaning~\cite{go86} as it is clear
from the previous section. In particular ~$Q$ plays the role of the exterior
derivative while ${\bar Q}$~ is the corresponding operation on vector fields.
The reason they are conserved is because the exterior derivative
anticommutes with the Lie-derivative.
If we interpret ~${\widetilde{\cal L}}$~ as a 1-D field theory, then
$Q$,${\bar Q}$~ are also analog of BRS-antiBRS charges\cite{be76} because their
graded $epb$-brackets are:~$\{Q,{\bar Q}\}_{epb}=0$,$\{Q,Q\}_{epb}=0$,
$\{{\bar Q},{\bar Q}\}_{epb}=0$. Besides these BRS-charges, there 
are also other charges conserved under ~${\widetilde{\cal H}}$
which make up, once combined with the ~$Q$~and~${\bar Q}$,
a genuine supersymmetry. They are~\cite{go90} ~
$N\equiv c^{a}\partial_{a}H$, ${\bar N}\equiv {\bar c}_{a}\omega^{ab}
\partial_{b}H$. The brackets among themselves and with~
$Q$~and~${\bar Q}$ are all zero except the following ones~$
\{Q,{\bar N}\}_{epb}=\sH~~~~,~~~\{i{\bar Q},iN\}_{epb}=\sH$.\linebreak
This is a N=4 supersymmetry. In fact these charges can be combined in the
following four independent ones:~$Q_{(1)}\equiv 
\frac{i(Q-{\bar N})}{\sqrt{2}}$,~$Q_{(2)}\equiv 
\frac{(Q+{\bar N})}{\sqrt{2}}$, \linebreak $Q_{(3)}\equiv 
\frac{i({\bar Q}+ N)}{\sqrt{2}}$,~$Q_{(4)}\equiv 
\frac{({\bar Q}-N)}{\sqrt{2}}$, all of which give ~$\{Q_{(i)}, 
Q_{(i)}\}_{epb}=\sH$\linebreak with~$(i)=(1)\cdots (4)$.
This {\it universal} supersymmetry of CM has a nice {\it physical}
interpretation deeply related to the fact that we are describing
a {\it classical} system. It basically says~\cite{go90} that the spontaneous
breaking of the invariance generated by the linear combinations
of some of the ~$Q_{(i)}$~above occurs for non-ergodic systems
while ergodic systems have that susy (at constant energy) always
unbroken.

We would like here to explore instead the {\it geometrical} meaning of
the susy charges . The reason is that all charges we found
previously had a nice geometrical interpretation, so also the
~$Q_{(i)}$ must have one. To better understand that, a first question  
we asked  ourselves is why God, in creating
such a fundamental object as the Lie-derivative ~$\sH$, did not think 
of making the Susy local. To a mathematician this way of studying the
geometrical meaning of a charge may sound strange, but not to a physicist.
So the first thing we decided to do is to go from the Lie-derivative
~$\sH$~(or its lagrangian~$\sL$), to a new one~$\sH_{g}$~(or~$\sL_{g}$)
which has that 
susy local. First (using the standard Noether technique~\cite{sugra74})
we do four {\it local} variations 
of ~$\sL$~generated by each of the four charges,
$Q$,${\bar Q}$,$N$,${\bar N}$, which make up the four supersymmetric
charges~$Q_{(i)}$  and test which one of these variations just produces
the associated charge itself multiplied by the time derivative of the 
symmetry parameter. The result~\cite{go98} is that only the variation under
~$Q$~and~${\bar N}$~do this. So these are the only two symmetry
generators which can be made local in the sense that, by introducing
auxiliary gauge-fields in~$\sL$,~the local variation of ~$\sL$~ can be
absorbed by the variation of the gauge-fields. As only two of the four
charges can be made local, we can say that from global N=4  we go down to a
local N=2 susy. In term of the susy charges~$Q_{(i)}$~ those which can be
made local are only~$Q_{(1)}$~and ~$Q_{(2)}$. The next step is to obtain
the expression of the local lagrangian~$\sL_{g}$. It is easy~\cite{go98} 
to find that: calling the "gauge" fields  ~$e,\psi_{1},\psi_{2}$~
where ~$e$ is a c-number field while ~$\psi_{1},\psi_{2}$ are Grassmannian 
ones, the ~$\sL_{g}$~is\linebreak $\sL_{g}\equiv\lambda_{a}{\dot \phi}^{a}+i{\bar c}_{a}
{\dot c}^{a}-e\sH+\psi_{1}{\bar N}+\psi_{2}Q$. It is not the first time that
a susy or a BRS has been made local in a 1-D system~\cite{div76}\cite{alv84}
\cite{za90} but never for
this action. The closest model for which it has been made local~\cite{alv84} 
is the supersymmetric quantum mechanics of Witten~\cite{wi81}(susyqm). The question now
to ask is if ~$\sL_{g}$~has any geometrical meaning at all. To do that we
have to explore the last three pieces present in ~$\sL_{g}$~ which were not
present in~$\sL$. The author of ref.\cite{alv84} indicates that there is a
gauge transformation which would turn the field ~$e$ into the constant 1 
while turning to zero both ~$\psi_{1}$~ and~$\psi_{2}$. In that manner
the extra piece would disappear and ~$\sL_{g}$~would turn into
$\sL$, at least for the  susyqm. This fact does not happen for sure
in our~$\sL_{g}$ as we will show now and we have {\it strong doubts} it happens 
even in susyqm~\cite{alv84}. The transformations of the gauge fields ~$e$,$\psi_{1}$,
$\psi_{2}$ under the two local transformations  associated
to~$\bar N$~and~$Q$~(with parameters~$\epsilon_{1}$, and ~$\epsilon_{2}$)
are respectively~$\{\delta_{1}\psi_{1}={\dot \epsilon}_{1},~\delta_{1}\psi_{2}
=0,~\delta_{1} e=\epsilon_{1}\psi_{2}\}$~and $\{\delta_{2}\psi_{2}=
{\dot \epsilon}_{2},~\delta_{2}\psi_{1}=0,~\delta_{2} e=-\epsilon_{2}
\psi_{1}\}$. As we have to bring, following~\cite{alv84},
~$\psi_{1}$~and~$\psi_{2}$~to zero, the infinitesimal
parameters which  do that are
$\epsilon_{1}(t)=-\int_{0}^{t}\psi_{1}(t^{\prime})dt^{\prime}$~and
$\epsilon_{2}(t)=-\int_{0}^{t}\psi_{1}(t^{\prime})dt^{\prime}$. Using these
parameters we have now at the same time to bring~$e$~to zero. The combined
transformation on ~$e$ would be:~$\delta e=[\int_{o}^{t}\psi_{2}dt^{\prime}]
\psi_{2}+[\int_{0}^{t}\psi_{1}dt^{\prime}]\psi_{1}$. This should be equal to
~$1-e$ to bring ~$e$ to zero but this is impossible because it would
impose a relation between~$e,\psi_{1},\psi_{2}$~ which does not exist.
So this fact gives us the information that it is not possible to
turn~$\sL_{g}$~into~$\sL$. The fields $e,\psi_{1},\psi_{2}$ remain, they have
no dynamic but they remain. They are the analog of the gravitons and gravitinos
in 1-D. They are basically Lagrangian multipliers
and the fact that they cannot be eliminated indicates that they have a role.
In fact if we insert ~$\sL_{g}$ into (1) we see that, once 
$e,\psi_{1},\psi_{2}$ are integrated out,
they impose some Dirac deltas constraints on the states between which
we sandwich the propagator. This selection of states is related to
the equivariant cohomology business~\cite{sto89} and this should bring
light to the geometrical meaning of ~$\sL_{g}$. More details will
appear in~\cite{go98}.

\end{document}